\documentclass{svproc}
\usepackage{url}

\usepackage{graphicx}        % standard LaTeX graphics tool when including figure files

\usepackage{multicol}        % used for the two-column index
\usepackage[bottom]{footmisc}% places footnotes at page bottom
\usepackage[label font = bf, position = top]{subfig} %labelformat
\usepackage{amsfonts}
\usepackage[cmex10]{amsmath}
\usepackage{float}
\usepackage{cite}
\usepackage{url}
\usepackage{ amssymb }
\usepackage{xcolor}
\usepackage{caption}
\usepackage{array}
\usepackage[section]{placeins}
\usepackage[utf8]{inputenc} % allow utf-8 input
\usepackage[T1]{fontenc}    % use 8-bit T1 fonts
\usepackage{hyperref}       % hyperlinks
\usepackage{url}            % simple URL typesetting
\usepackage{booktabs}       % professional-quality tables
\usepackage{amsfonts}       % blackboard math symbols
\usepackage{nicefrac}       % compact symbols for 1/2, etc.
\usepackage{microtype}      % microtypography
\usepackage{lipsum}

\begin{document}
\mainmatter              % start of a contribution
\title{Sector Neutral Portfolios: Long memory motifs persistence in market structure dynamics}
\titlerunning{Long memory motifs}  % abbreviated title (for running head)
%                                     also used for the TOC unless
%                                     \toctitle is used
%
\author{Jeremy D. Turiel\inst{1} \and Tomaso Aste\inst{1, 2, 3}}
\authorrunning{Jeremy Turiel and Tomaso Aste} % abbreviated author list (for running head)
%
%%%% list of authors for the TOC (use if author list has to be modified)
\tocauthor{Jeremy D. Turiel, Tomaso Aste}
\institute{Department of Computer Science, University College London, Gower St, Bloomsbury, London WC1E 6BT, United Kingdom,\\
\email{jeremy.turiel.18@ucl.ac.uk}, \email{t.aste@ucl.ac.uk},\\ WWW home page:
\texttt{http://www.cs.ucl.ac.uk/staff/tomaso$\_$aste/}
\and
UCL Centre for Blockchain Technologies, University College London, Gower St, Bloomsbury, London WC1E 6BT, United Kingdom \\ WWW home page: \texttt{http://blockchain.cs.ucl.ac.uk/tomaso-aste/}
\and
Systemic Risk Centre, London School of Economics and Political Sciences, Houghton Street, London WC2A 2AE, UK}

\maketitle              % typeset the title of the contribution

\begin{abstract}
We study soft persistence (existence in subsequent temporal layers of motifs from the initial layer) of motif structures in Triangulated Maximally Filtered Graphs (TMFG) generated from time-varying Kendall correlation matrices computed from stock prices log-returns over rolling windows with exponential smoothing. We observe long-memory processes in these structures in the form of power law decays in the number of persistent motifs. The decays then transition to a plateau regime with a power-law decay with smaller exponent. We demonstrate that identifying persistent motifs allows for forecasting and applications to portfolio diversification.
Balanced portfolios are often constructed from the analysis of historic correlations, however not all past correlations are persistently reflected into the future. Sector neutrality has also been a central theme in portfolio diversification and systemic risk. We present an unsupervised technique to identify persistently correlated sets of stocks. These are empirically found to identify sectors driven by strong fundamentals. Applications of these findings are tested in two distinct ways on four different markets, resulting in significant reduction in portfolio volatility. A persistence-based measure for portfolio allocation is proposed and shown to outperform volatility weighting when tested out of sample.
%``Soft'' persistence of motif structures (existence in subsequent temporal layers of motifs from the initial layer) is then studied.
% We would like to encourage you to list your keywords within
% the abstract section using the \keywords{...} command.
\end{abstract}

\section{Introduction} \label{intro}

%Correlation networks \cite{mantegna1999hierarchical} applied to financial assets have recently gained wide attention, together with associated network filtering techniques, as these methods show that meaningful taxonomy of financial assets is identifiable from these sparse network structures. Filtering through the MST technique was initially suggested by Mantegna \cite{mantegna1999hierarchical}, while this concept was taken further in \cite{massara2016network} \cite{tumminello2005tool}, allowing for chordal graphs with a predefined motif structure such as the PMFG.

Portfolio diversification has been a central theme for investors and the financial industry since the beginning of the $20^{th}$ century, with the first long-lasting academic results produced by Harry Markowiz \cite{markowitz1952portfolio}. Sector diversification, in particular, is a non-trivial type of diversification for localised systemic risk. This risk arises from persistently correlated groups of stocks which often correspond to industry sectors. Correlations within these groups of stocks are found to be highly persistent in time, this should be accounted for when allocating capital within a portfolio. %Solid results in applications to portfolio diversification are presented in Section \ref{results}. 
These persistently correlated groups are often subject to similar regulatory, political and resource shocks. Examples of this are stocks belonging to the Oil and Gas, Healthcare or Pharmaceutical sectors. It will be shown in the present work that (absolute) correlation values and their persistence should be treated differently for the purpose of portfolio diversification as they represent different properties of the system and are only weakly related.

%After the financial crisis, market structures changed and sectors became less clearly defined by the underlying market structure \cite{musmeci2014risk}. Stimulated by this discussion,
The present work introduces an unsupervised technique to identify groups of stocks which share strong fundamental price drivers. This technique can be of particular impact in less traded markets, where identifying structures with shared fundamental price drivers might require in-depth knowledge of the companies. A persistence-based measure is also proposed to optimise portfolio allocation and tested for out of sample performance against $1/\sigma$ weighting (where $\sigma$ is the standard deviation of log-returns from which the correlation matrix is constructed).
%In the present work we therefore provide an unsupervised learning technique to identify persistent market structures, with solid results in its applications to portfolio diversification.

Correlation networks \cite{mantegna1999hierarchical} and network filtering techniques applied to the study of financial assets have recently gained wide attention \cite{marcaccioli2019polya, massara2019learning, micciche2019primer, musciotto2018bootstrap, jovanovic2018financial, cimini2019statistical, kojaku2019constructing, masuda2018configuration}. These methods show that meaningful taxonomy of financial assets is identifiable from these sparse network structures. Filtering through the Minimum Spanning Tree (MST) technique was initially suggested by Mantegna \cite{mantegna1999hierarchical}, this concept was further extended to planar graphs with the Planar Maximally Filtered Graph (PMFG) \cite{tumminello2005tool} and more recently to chordal graphs with predefined motif structure, as the TMFG in \cite{massara2016network} and the Maximally Filtered Clique Forest (MCFC) in \cite{massara2019learning}.
%Recently Hosseini et al. \cite{hosseini2019weight} have suggested a new filtering technique based on normalised mutual information between stocks and directly compared it with the PMFG \cite{tumminello2005tool}.

Correlations are noisy measures of comovement of financial asset prices, which are often non-stationary within the observation window. Longer windows benefit the measure's stability, as we have more observations to estimate the $N (N-1)/2$ parameters of the matrix of $N$ assets. However, a longer observation window can come with the disadvantage of weighting more and less recent co-movements equally with the risk of averaging over multiple non-stationarities. In order to compensate for this effect, we apply the exponential smoothing method as discussed in Pozzi et al. for Kendall correlations \cite{pozzi2012exponential}. This allows for more stable correlations, while prioritising recent observations. The method applies an exponential weighting to the correlation window, prioritising more recently observed comovements. %Here we use Kendall correlations with exponential smoothing as per \cite{pozzi2012exponential}.

The rest of the paper is structured as follows: Section \ref{method} describes the methods applied and defines measures which are used throughout the paper. Section \ref{results} describes the results obtained, with Section \ref{results-long-term} introducing long-term memory processes in persistence, Section \ref{class_decay_exp} analysing market development through its decay exponents, Section \ref{sector_motifs} illustrating the coherence of highly persistent motifs with sectors and Section \ref{portfolio_applications} outlining various results which highlight the importance of this work for portfolio allocation. Section \ref{analysis} then presents an analysis of the results from Section \ref{results} and Section \ref{conclusion} concludes the paper with a summary and thoughts for further work.

%due to their applications to the characterisation and modelling of socio-economic systems.

\section{Method}\label{method}

%We select up to 500 most capitalised stocks (at the moment of data collection) for the following markets: NYSE, Germany, Italy, China (Shenzen), Israel, Russia, Greece, Turkey, United Kingdom, Saudi Arabia, Mexico, India (Bombay), Japan (Tokio), France (Paris Euronext), Venezuela.

In the present paper we apply the TMFG \cite{massara2016network} to filter matrices obtained from Kendall correlations with exponential smoothing, applying the method by Pozzi et al. \cite{pozzi2012exponential}.

\subsection{Data} \label{data}

We select the 100 most capitalised stocks from the NYSE, Italy, Germany and Israel's markets (400 in total). Markets range from highly liquid and more developed ones such as the NYSE and Germany to less liquid and stable markets such as Italy and Israel.

We investigate daily closing price data from Bloomberg between 3/01/2014 for the NYSE, Germany and Italy (5/01/2014 for Israel) and 31/12/2018 (inclusive) for the NYSE (28/12/2018 for Germany and Italy, 1/1/2019 for Israel). The data is composed of 1258 observations for the NYSE, 1272 for Italy and Germany and 1225 for Israel.
%We though notice that some markets do not have sufficient/good quality data easily available, both in terms of number of stocks and price history. These are hence left for further work and we restrict ourselves to the abovementioned markets. 100 most capitalised stocks (at the moment of data collection) for the following markets: NYSE, Germany, Italy, Israel.

\subsection{TMFG network motif persistence}

We look at temporal persistence of tetrahedral and triangular motifs in the TMFGs constructed over rolling windows. TMFG networks can be viewed as trees of tetrahedral (maximal) cliques connected by common triangular faces, these are then triangular cliques with different meaning in the taxonomy, called separators. Not all triangular faces of the tetrahedral cliques are separators and we will refer to those which are not as triangles (these do not include separators in the way we shall refer to them).

Differently from ``hard'' persistence (survival) of motifs between consecutive layers in the temporal network, which is more common in the literature \cite{dessi2018supernoder, musmeci2014risk}, here we apply a form of ``soft'' persistence. A motif corresponding to clique $\mathcal{X}_c$ is considered ``soft'' persistent at time $t + \tau$ if and only if the motif is present at both the initial time $t$ and at $t + \tau$. %This can be formalised as:

Considering the motif sets $\mathcal{X}_C^{t} = \{ \mathcal{X}_i^{t} \}_{i = 1, ..., C}$ and $\mathcal{X}_C^{t + \tau} = \{ \mathcal{X}_i^{t + \tau} \}_{i = 1, ..., C}$, the binary persistence value of motif $c \in C$ at time $t$ and $t + \tau$ is

\begin{equation}
\centering
P_{m}(\mathcal{X}_c^{t, t+\tau}) = (\mathcal{X}_c \in \mathcal{X}_C^{t}) \land (\mathcal{X}_c \in \mathcal{X}_C^{t + \tau})
\label{eq_defin_persist}
\end{equation}

Where $P_{m}(\mathcal{X}_c^{t, t+\tau})$ represents the binary persistence value of motif $c \in C$ at times $t$ and $t + \tau$.

%in the TMFG network obtained from the correlation matrix (Kendall with exponential smoothing)

\subsection{Portfolio construction}

We investigate the decay in the number of persistent motifs between filtered TMFG correlation networks with observation windows progressively shifted by one trading day. We iterate over $t = \left[ 0, ..., 200 \right[$ different starting correlation networks and investigate persistence up to a time shift of $\tau = 900$ days. Hence the analysis covers a significant portion of temporal layers which do not overlap with the time window of the initial layer.
We observe $\langle P_{m}(\mathcal{X}^{\tau}) \rangle_{T, C}$ from Equation \ref{eq_persist_motif_avg} to decay with the time shift $\tau$.
We then obtain the power law fit for the decay law and identify the two regimes: one with a faster decay followed by one with a slower decay. The transition point is computed by minimising the unweighted average mean squared error (MSE) between the two fits over all possible transition points in time.

The average motif persistence in the plateau regime is defined as

\begin{equation}
\centering
\langle P_{m}(\mathcal{X}_c) \rangle_{T, \mathcal{T}} = \frac{1}{T} \cdot \frac{1}{\mathcal{T} - \tau_{plat}} \cdot \sum_{t = 0}^{T} \sum_{\tau = \tau_{plat}}^{\mathcal{T}} P_{m}(\mathcal{X}_c^{t, t+\tau})
\label{eq_persist_time_avg}
\end{equation}

Where $\tau_{plat}$ identifies the transition point to the plateau region identified by minimising the Mean Squared Error (MSE), as explained below.

The average persistence for the entire clique set over $T$ starting points at time shift $\tau$ is defined as

\begin{equation}
\centering
\langle P_{m}(\mathcal{X}^{\tau}) \rangle_{T, C} = \frac{1}{T} \cdot \frac{1}{|C|} \cdot \sum_{t = 0}^{T} \sum_{c \in C} P_{m}(\mathcal{X}_c^{t, t+\tau})
\label{eq_persist_motif_avg}
\end{equation}

We also compare the decay exponents for multiple random stock selections over different markets to identify whether the steepness of motif decay (edge, closed triad or tetrahedron clique) is indicative of market stability/development stage.
We further investigate more liquid markets such as the NYSE from both a quantitative and qualitative point of view.% as follows.
We classify motifs in the plateau by their ``soft'' persistence and study the sector structure of the most persistent motifs. We also verify that these motifs are not trivially retrieved by maximum correlation edges or motifs in the correlation matrix.

In order to further justify the analysis of motifs over individual edges, we test to reject the assumption that motifs are formed by edges in the network whose existence is not mutually dependent. The assumption would imply that coexistence of edges in motifs is not statistically significant and that motif structures have no extra persistence beyond the individual edges that form them. The hypothesis being tested implies that motif persistence is simply the result of persistence characterising their component edges:
%This is falsified by the consistently lower decay exponent (in modulus) for adjusted persistence of triangular motifs.

\begin{equation}
\centering
P_{m}(\boldsymbol{\chi}_c^{t,t+ \tau}) = P_{m}(\boldsymbol{\chi}_{c1}^{t,t + \tau}) \cdot P_{m}(\boldsymbol{\chi}_{c1}^{t,t + \tau}) \cdot P_{m}(\boldsymbol{\chi}_{c3}^{t,t + \tau})
\label{eq_motif_meaning}
\end{equation}

Where the motif and its edges are defined as $\boldsymbol{\chi}_c^{t,t + \tau} = \{ \boldsymbol{\chi}_{c1}^{t,t + \tau} , \boldsymbol{\chi}_{c2}^{t,t + \tau} , \boldsymbol{\chi}_{c3}^{t,t + \tau} \}$.

In order to provide an industry-oriented point of view, we construct a portfolio containing all stocks in the ten most persistent motifs (for each market) and compare its volatility with that of random portfolios. %We also construct a portfolio containing all stocks in the ten most persistent motifs and compare its volatility with that of portfolios containing only part of the motifs (breaking the persistent motif structure).

We conclude by defining the persistence measure $P_m(v_i)$ in Equation \ref{eq_portfolio_measure} to compare random portfolios weighted by $1/ \sigma$ with those weighted by $1/ P_m(v_i)$. We do this for the four different markets, with all results showing meaningful volatility reductions.

\begin{equation}
\centering
P_m(v_i) = \sum_{\mathcal{X}_c \in \mathcal{X}_C | i \in \mathcal{X}_c} \langle P_{m}(\mathcal{X}_c) \rangle_{T, \mathcal{T}}
\label{eq_portfolio_measure}
\end{equation}

The measure presented in Equation \ref{eq_portfolio_measure} is defined for each vertex $v_i$ in the network as the sum over all $\langle P_{m}(\mathcal{X}_c) \rangle_{T, \mathcal{T}}$ (average pesistence of motif $\mathcal{X}_c$ in the plateau) where vertex $v_i$ belongs to clique $\mathcal{X}_c$. %, with  the vertex of the network associated with the measure.

\section{Results} \label{results}

\subsection{Long-term memory of motif structures} \label{results-long-term}

\begin{figure}[!htb]
\centering
\includegraphics[width=1\textwidth]{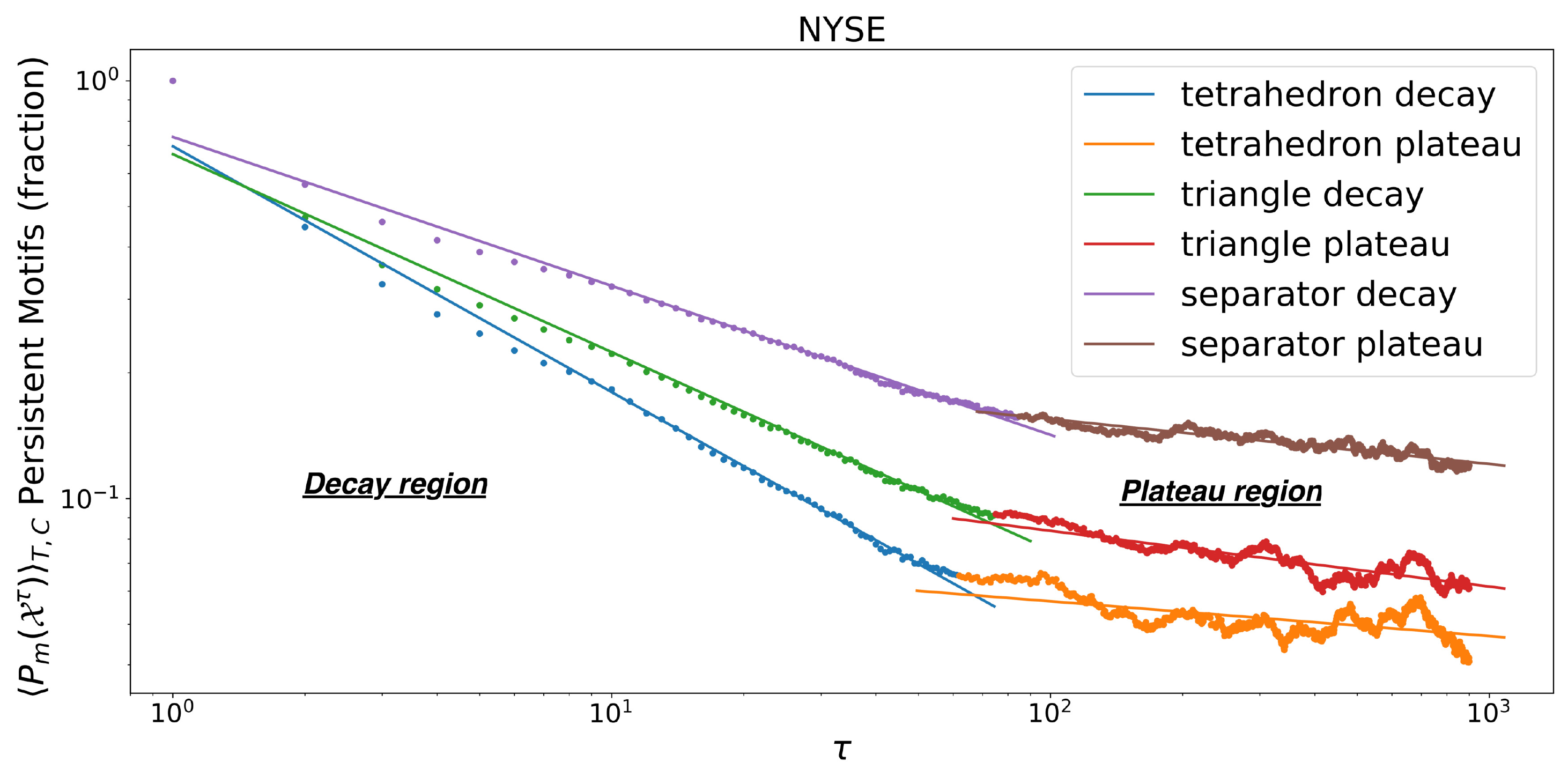}
\caption{Decay of triangular clique faces, separators and clique motifs overlap between layers for 100 NYSE stocks, as a function of time interval $\delta_t = [0, 900]$ (average over 200 simulations). The two power-law regimes are identified by the minimum MSE sum of the fits.}
\label{figure2}       % Give a unique label
%\vspace{-0.4cm}
\end{figure}

%\begin{figure}[!htb]
%\centering
%\includegraphics[width=1\textwidth]{power_law_decay_NYSE.pdf}
%\caption{Decay of triangular clique faces, separators and clique motifs overlap between layers for 100 NYSE stocks, as a function of time shift $\tau = [0, 900]$ (average over 200 values of $t$). The two power-law regimes are identified by the minimum MSE sum of the fits.}
%\label{figure1}       % Give a unique label
%\vspace{-0.4cm}
%\end{figure}
%in the number of ``soft'' persistent motifs
The plot in Figure \ref{figure2} shows the power law decay (evident from the linear trend in log-log scale) in $\langle P_{m}(\mathcal{X}^{\tau}) \rangle_{T = 200, C}$ vs. $\tau$, followed by a plateau region. We also observe that all motif decays have $\tau_{plat} = [\delta t_{window}/2, \delta t_{window}]$, where $\delta t_{window}$ is the initial window's time span. The window used has $\delta t_{window} = 126$ trading days and a value of $\theta = 46$ for exponential smoothing, as per \cite{pozzi2012exponential}. The choice of $\delta t_{window}$ corresponds to roughly 6 months of trading and satisfies $N < \delta t_{window}$, with $N$ the number of assets in the correlation matrix. The correlation matrix is hence well-conditioned and invertible.

As per the plot in Figure \ref{figure2}, there are $N - 3 = 97$ cliques in the starting TMFG networks and $3N - 8 = 292$ face triangles.

%The plot in Figure \ref{figure2} is analogous to that in Figure \ref{figure1}, but with the number of persistent motifs scaled by the total number of motifs in an $N$-nodes TMFG. This corresponds to the number of motifs in the first temporal layer.
%Figure \ref{figure2} provides a better visual comparison of decay rates, with tetrahedral cliques having a higher decay rate, as they are composed of more edges (six) which need to persist simultaneously.

In Figure \ref{figure2} we notice that the minimum MSE for the two linear fits is achieved at the transition point between the decay phase and the plateau. The transition point can therefore be identified by minimising a standard fit measure, this strengthens the unsupervised nature of our method. The method for minimum MSE search is described in Section \ref{method}.

\subsection{Market classification via decay exponent} \label{class_decay_exp}

We now consider how the decay exponent behaves across markets. Table \ref{tab1} compares the decay exponents for cliques, triangular motifs and clique separators in the NYSE, German stock market, Italian stock market and Israeli stock market.

\begin{table}
\caption{Exponents for the decay power law regime computed with MSE. The analysis refers to 100 randomly selected stocks amongst the 500 most capitalised, over time intervals $\tau = \left[0, 900 \right[$ and $t = \left[ 0, ..., 200 \right[$ different initial temporal network layers. For all motif analyses in this work triangles and separators constitute non-overlapping sets, as these represent theoretically and taxonomically different structures and decay characteristics.}
\centering
  \begin{tabular}{llll}
    \toprule
    %\multicolumn{2}{c}{Part}                   \\
    %\cmidrule(r){1-2}
Market & Clique & Triangular Motif & Clique Separator \\
\midrule
NYSE & -0.392 & -0.493  & -0.245 \\
Germany & -0.792 & -0.598  & -0.381  \\
Italy & -0.785 & -0.811  & -0.174*  \\
Israel & -1.024 & -0.866  & -0.728  \\
    \bottomrule
  \end{tabular}
  \footnotesize

* Result compromised by regimes not well identified for motif decay in large systems ($\approx 100$ stocks)
  \label{tab1}
\end{table}

We notice from the results in Table \ref{tab1} that the NYSE, which is clearly the most developed and liquid stock market, has the lowest decay exponent (in modulus, which corresponds to the slowest decay) for both cliques and triangles. This indicates that its correlations are more stable on a shorter time window, due to a higher signal to noise ratio. Germany and Italy have similar values for clique exponents, with Germany seemingly more stable in terms of triangular motifs. Israel, a younger and less liquid stock market, follows with a faster decay in both cliques and triangular motifs. The ordering of these markets is not clearly identifiable in clique separators as noise in the data does not allow for the two decay regimes discussed in Section \ref{results-long-term} to be correctly identified in all markets (in this case for Italy). Separators have a distinct role and meaning in the graph's taxonomy and further work should allow for a more thorough analysis of those.

%Tetrahedral cliques can be viewed as groups of stocks representing maximal cliques (of four vertices) in the graph, while triangles are smaller cliques (three-cliques) which constitute the faces of the tetrahedral maximal cliques.
In Table \ref{tab1} the decay exponent is not adjusted by the probability that all edges in the clique must be present in the temporal layer for the clique to exist. We show in Table \ref{tab2} that, when adjusted by the probability of all its edges existing simultaneously, triangular motifs have a slower decay than individual edges. In order to strengthen the consistence of the phenomenon across buckets of randomly selected stocks, Table \ref{tab2} corresponds to a different random bucket than Table \ref{tab1}.

We stress that Table \ref{tab2} falsifies the hypothesis discussed in Section \ref{method} that motifs are formed by edges in the network whose existence is not mutually dependent. This is falsified by the consistently lower decay exponent (in modulus) for adjusted persistence of triangular motifs. We can then conclude that motifs are more stable structures across temporal layers of the network, with significant interdependencies in their edges' existence.

\begin{table}
\caption{Exponent for the power law decay regime identified by MSE in different sample markets. The analysis refers to 100 randomly selected stocks amongst the 500 most capitalised, over time intervals $\tau = \left[0, 900 \right[$ and $t = \left[ 0, ..., 200 \right[$ different initial temporal network layers.}  \centering
  \begin{tabular}{llll}
    \toprule
    %\multicolumn{2}{c}{Part}                   \\
    %\cmidrule(r){1-2}
Market & Edge & Triangular Motif & Triangular Motif** \\
\midrule
NYSE & -0.164 & -0.398 & -0.133 \\
Germany & -0.265 & -0.471 & -0.157 \\
Italy & -0.144* & -0.458 & -0.153 \\
Israel & -0.397  & -0.830 & -0.277 \\
    \bottomrule
  \end{tabular}
  \footnotesize

* Result compromised by regimes not well identified for edge decay in large systems ($\approx 100$ stocks)

** Motif exponent adjusted by the probability of simultaneous edge persistence in the motif
  \label{tab2}
\end{table}

\subsection{Sector analysis in persistent motifs} \label{sector_motifs}

\begin{figure}
\centering
\includegraphics[width=120mm]{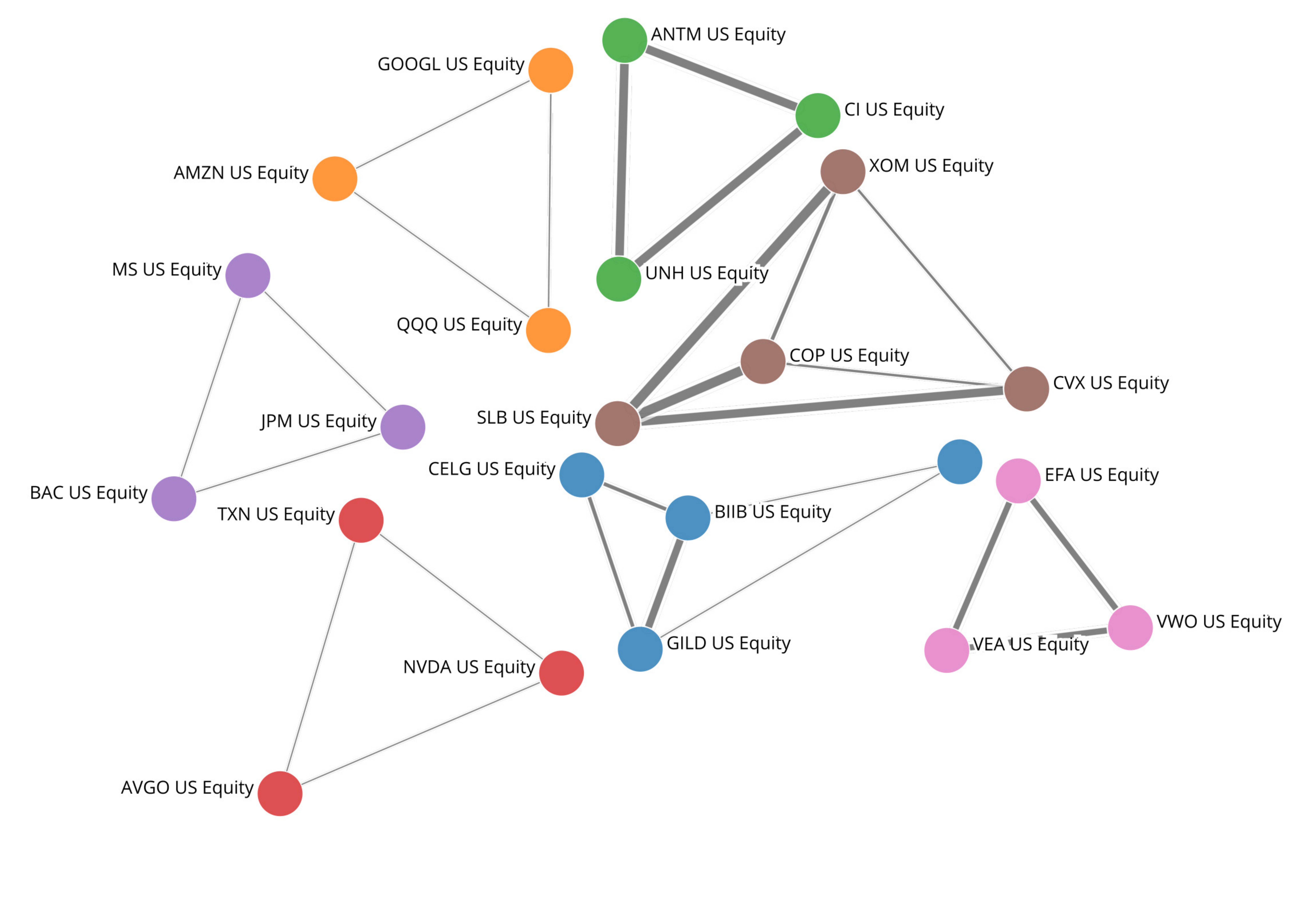}
\caption{Network representation of the ten most persistent triangular motifs in the TMFG layers for the 100 most capitalised stocks of the NYSE.}
\label{figure3}       % Give a unique label
%\vspace{-0.4cm}
\end{figure}

Figure \ref{figure3} provides a visualisation of the network components formed by the ten most persistent triangles in the NYSE. We observe that all strongly persistent triangles have elements which belong to the same industry sector. Table \ref{tab3} shows this for the ten most persistent triangles displayed in Figure \ref{figure3}.
%an illustration of this for the triangles in Figure \ref{figure3} is provided in Table \ref{tab3}.

We notice that most sectors for the motifs in Table \ref{tab3} share strong fundamental price drivers, which justify the persistent structure in the long term, as per the discussion in Section \ref{intro}. Other motifs are constituted by ETFs with similar underlying assets (Vanguard FTSE ETF, MSCI EAFE ETF, Vanguard FTSE ETF) or the NASDAQ ETF with its main holdings (Amazon and Alphabet). The reason for the existence of these motifs is intuitive and does not affect our analysis, as ETF-related motifs are unlikely to be present in the network formed by a random selection of stocks or by stocks in a portfolio. These motifs are present here as we focus on the 100 most capitalised securities in the NYSE, which include ETFs.

\begin{table}

 \caption{Motif components and Financial Times sector affiliation for the ten most persistent motifs in the NYSE's 100 most capitalised stocks.}
  \centering
  %\tiny
  %\setlength\tabcolsep{1pt}
  \begin{tabular}{llllll}
    \toprule
    %\multicolumn{2}{c}{Part}                   \\
    %\cmidrule(r){1-2}
    Security 1 & Security 2 & Security 3 & FT Sector \\
    
    \midrule

    Biogen Inc & Gilead Sciences Inc & Celgene Corp & Biopharmaceutical \\

    UnitedHealth Group Inc & Cigna Corp & Anthem Inc & Health Care \\

    Biogen Inc & Gilead Sciences Inc & Amgen Inc & Biopharma/tech \\

    Bank of America Corp & JPMorgan Chase \& Co & Morgan Stanley & Financials-Banks \\
    
    Vanguard FTSE ETF** & MSCI EAFE ETF & Vanguard FTSE ETF*** & Index ETFs \\
    
    Invesco QQQ Trust* & Amazon.com Inc & Alphabet Inc & Tech \\

    ConocoPhillips & Schlumberger NV & Exxon Mobil Corp & Oil \& Gas \\

    NVIDIA Corp & Texas Instruments Inc & Broadcom Inc & Tech Hardware \\

    Chevron Corp & Schlumberger NV & Exxon Mobil Corp & Oil \& Gas \\

    Chevron Corp & ConocoPhillips & Schlumberger NV & Oil \& Gas \\
    
    \bottomrule
  \end{tabular}
  \footnotesize

* ETF on NASDAQ - Top Holdings include Amazon, Facebook, Apple, Alphabet

** Vanguard FTSE Developed Markets Index Fund ETF Shares

*** Vanguard FTSE Emerging Markets Index Fund ETF Shares
  \label{tab3}
\end{table}

We also investigate whether motif persistence and motif structures can be easily retrieved from the original correlation matrix. The purpose of this is to check that our method is not redundant and trivially replaceable.
To test this, we consider the ten most present persistent triangles across the plateau region and check their overlap with the ten most correlated triplets in each unfiltered correlation matrix. We find that no more than one triangle lies in the intersection between the two sets, in each temporal layer. %corresponds in each layer.
We also check the correlation between motif persistence and the average sum or product (results are equivalent for our purpose) of its individual edges' correlation for all unfiltered correlation layers. We observed both visually (by means of different plots) and statistically (through the Pearson and Kendall correlation measures) that the two measures are only loosely related, if they are at all. This was reflected by plots with no clear trend or apparent functional form. The variables also presented significant, low correlation values, where the correlation explained no more than $20\%$ of the variance in the persistence values. This result is especially significant for a wide power law distribution as that of persistence values. %Figure \ref{figure_trivial_corr} shows a scatter plot demonstrating that the measure is not trivially retrievable from the correlation matrix.

\subsection{Long-only portfolio diversification across markets}\label{portfolio_applications}

\subsubsection{Motif vs. random portfolios}

\begin{figure}
    \centering
    %\ContinuedFloat
    %\setcounter{subfigure}{0}
    \subfloat[NYSE]{\includegraphics[width=0.49\linewidth]{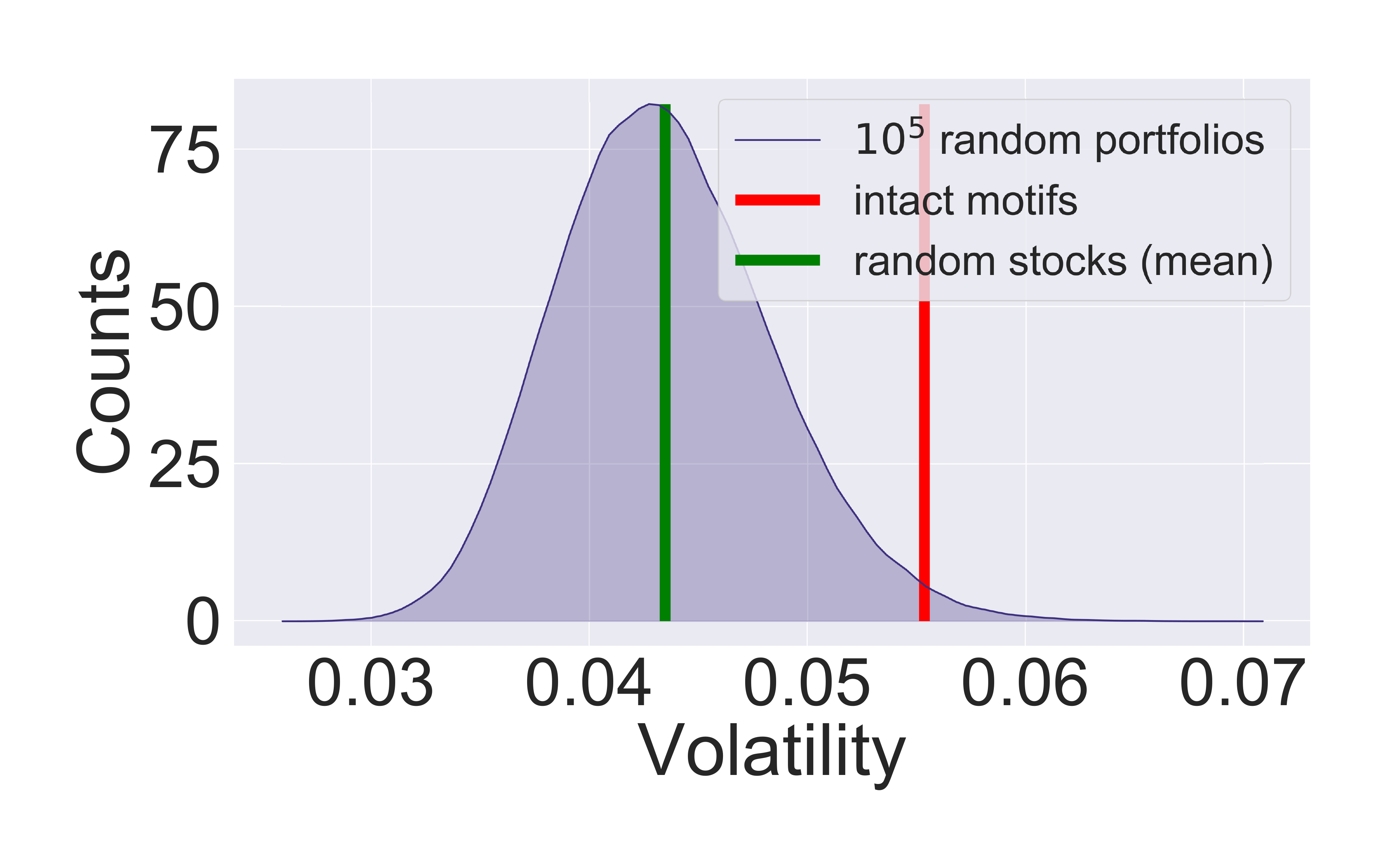}\label{figure4}}
    \subfloat[Germany]{\includegraphics[width=0.49\linewidth]{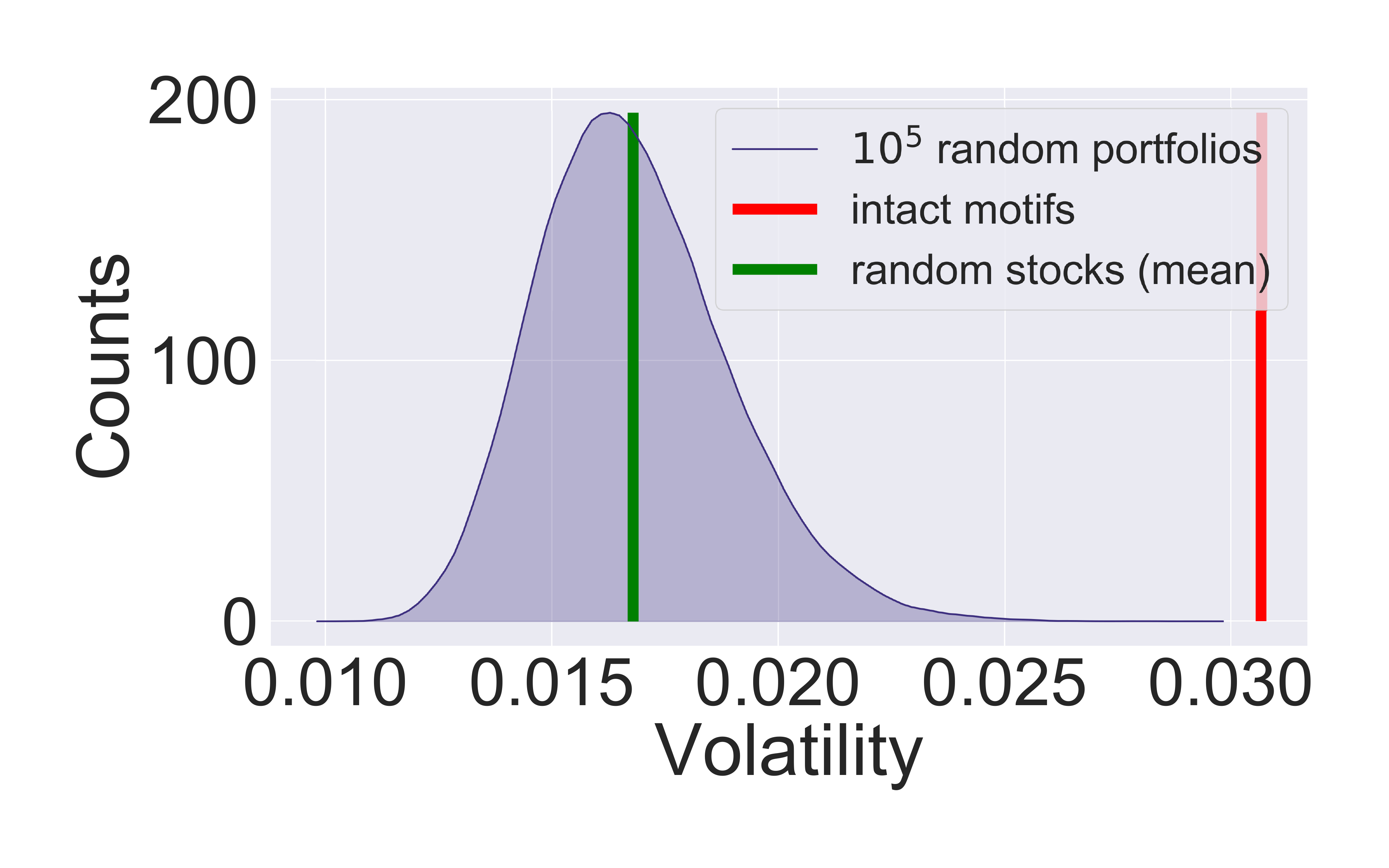}\label{figure5}}
    \\
    %\vspace{1cm}
    \subfloat[Italy]{\includegraphics[width=0.49\linewidth]{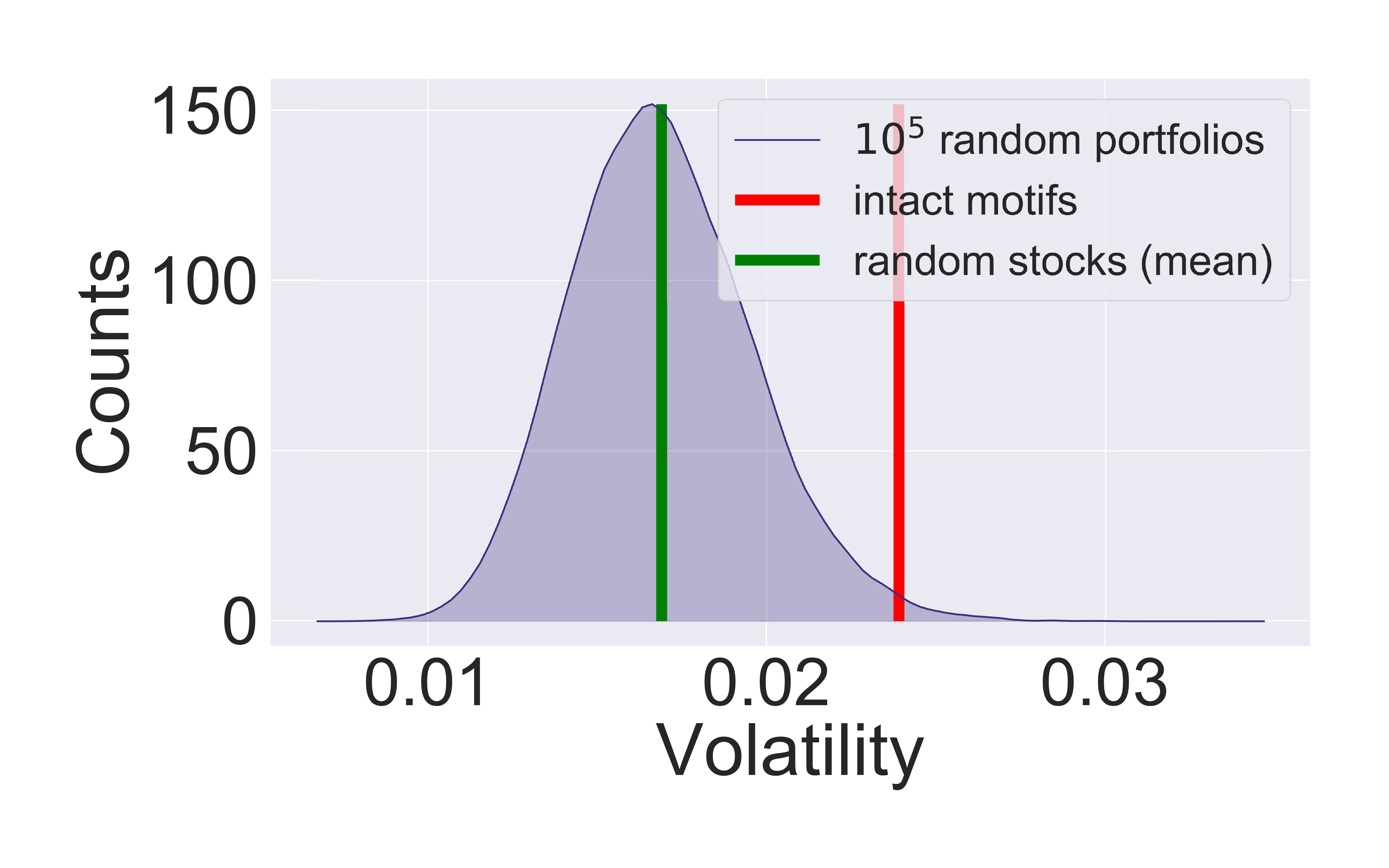}\label{figure6}}
    \subfloat[Israel]{\includegraphics[width=0.49\linewidth]{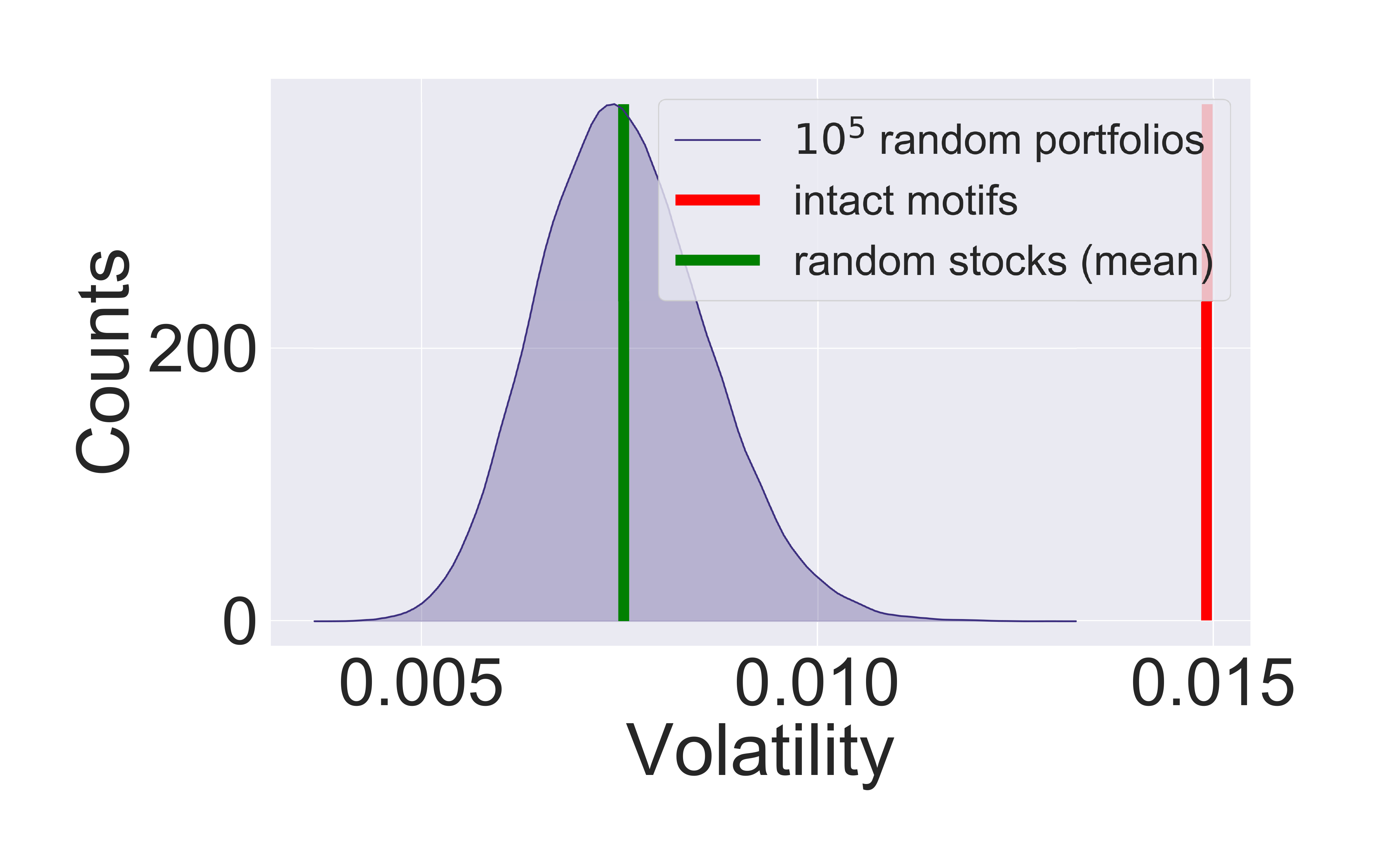}\label{figure7}}
    \caption{Portfolio volatility distribution for the 100 most capitalised stocks in the NYSE (a), German stock market (b), Italian stock market (c) and Israeli stock market (d). The reference portfolio (red bar) contains all stocks in the 10 most persistent triangles and distribution portfolios are formed from a random selection of stocks (mean distribution volatility represented by the green bar).}
    \label{fig_random_portfolios}
\end{figure}

We check that a portfolio formed by the 10 most persistent motifs in each market has a highly enhanced out of sample volatility due to its stable correlations.

This is shown in Figure \ref{fig_random_portfolios} where we consider the volatility of the motif portfolio and a distribution of volatilities for $10^5$ randomly selected portfolios with the same number of stocks.

As expected, we observe the motif portfolio to yield a volatility close to the higher end of the distribution. We should highlight that the volatility of portfolios is evaluated out of sample with respect to the period the persistence was calculated on, making this method not only observational, but also predictive.

\subsubsection{Volatility vs. persistence weighting - an industry standard comparison}

\begin{figure}[H]
    \centering
    %\ContinuedFloat
    %\setcounter{subfigure}{0}
    \subfloat[NYSE]{\includegraphics[width=0.49\linewidth]{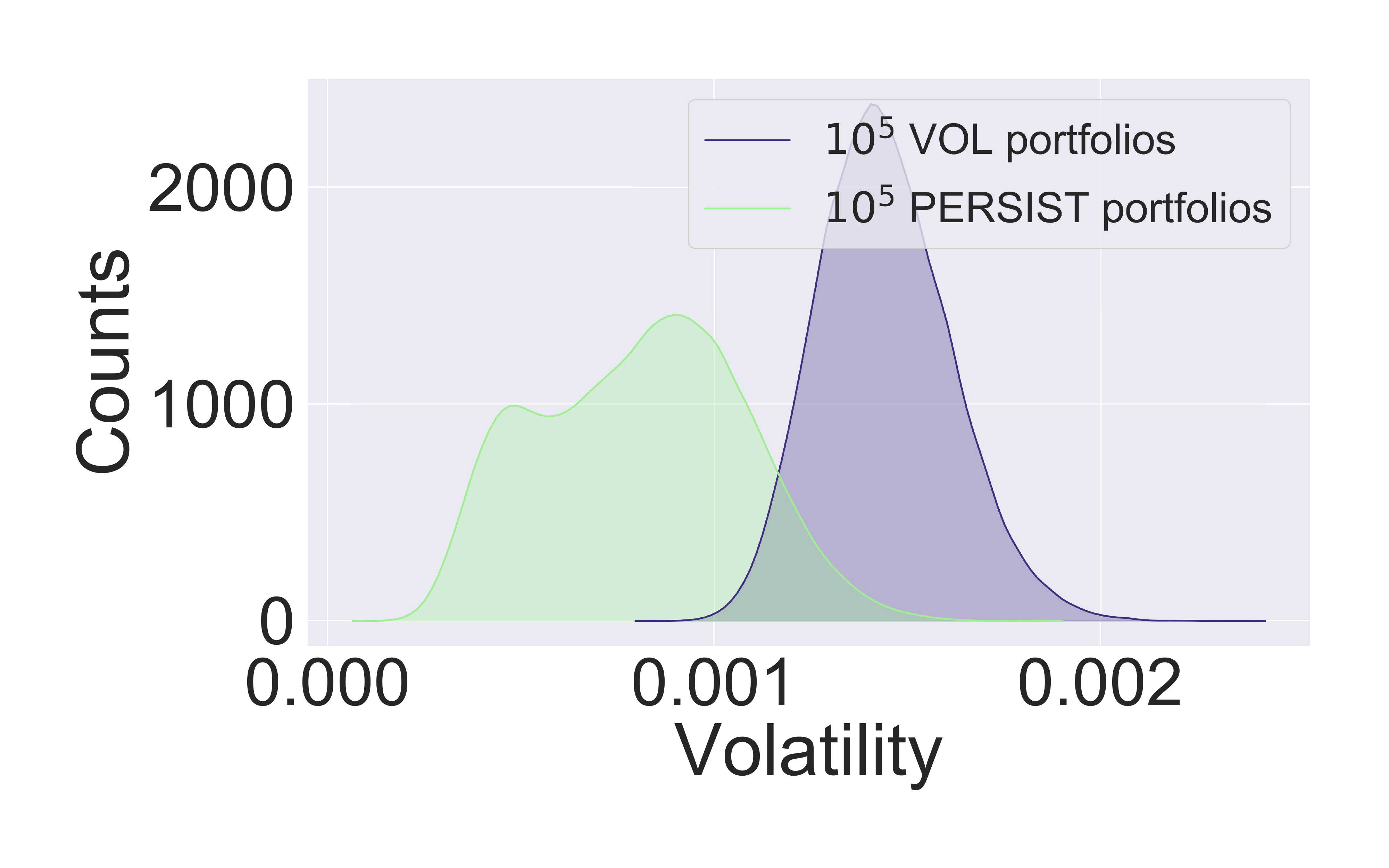}\label{figure8}}
    \subfloat[Germany]{\includegraphics[width=0.49\linewidth]{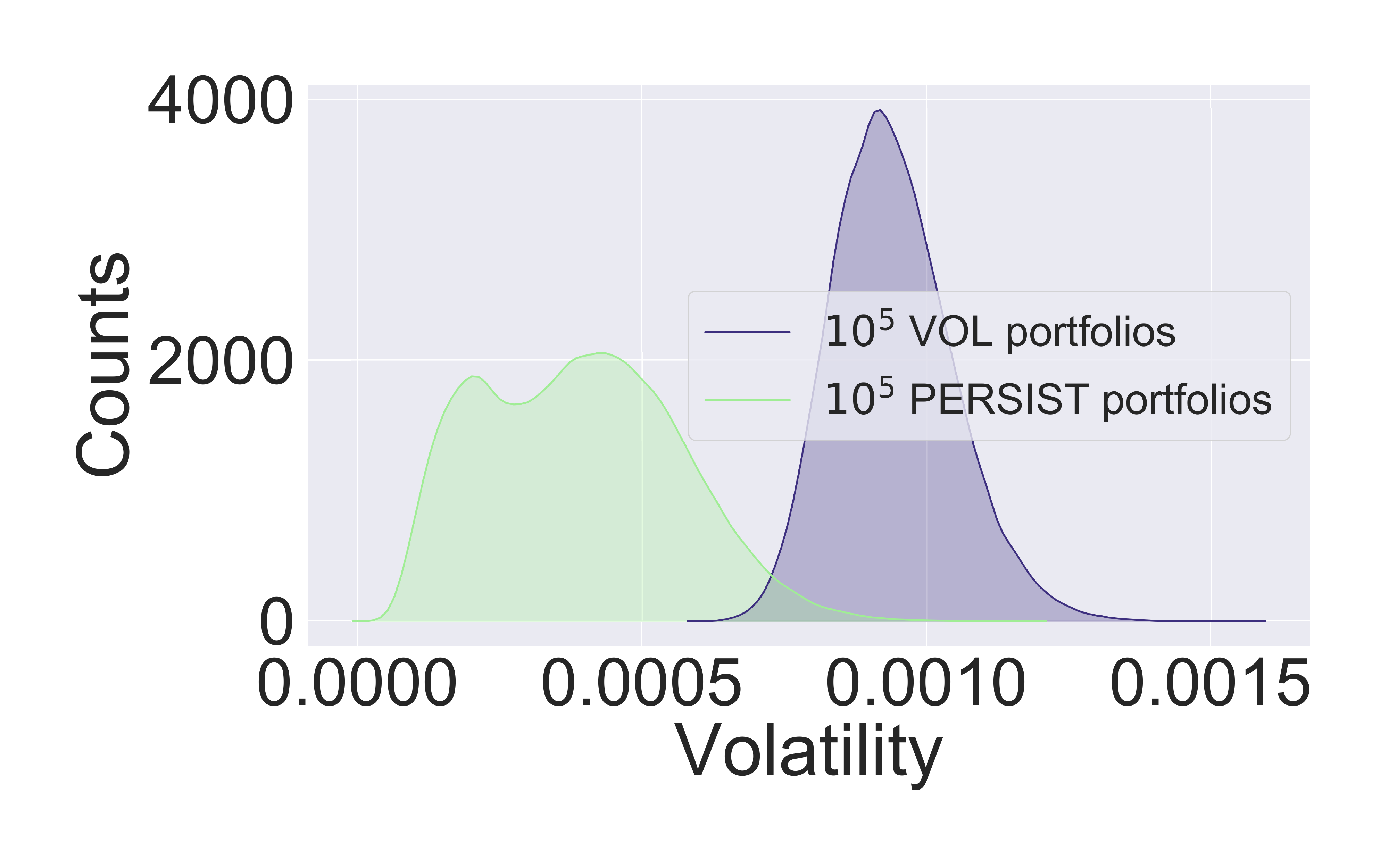}\label{figure9}}
    \\
    %\vspace{1cm}
    \subfloat[Italy]{\includegraphics[width=0.49\linewidth]{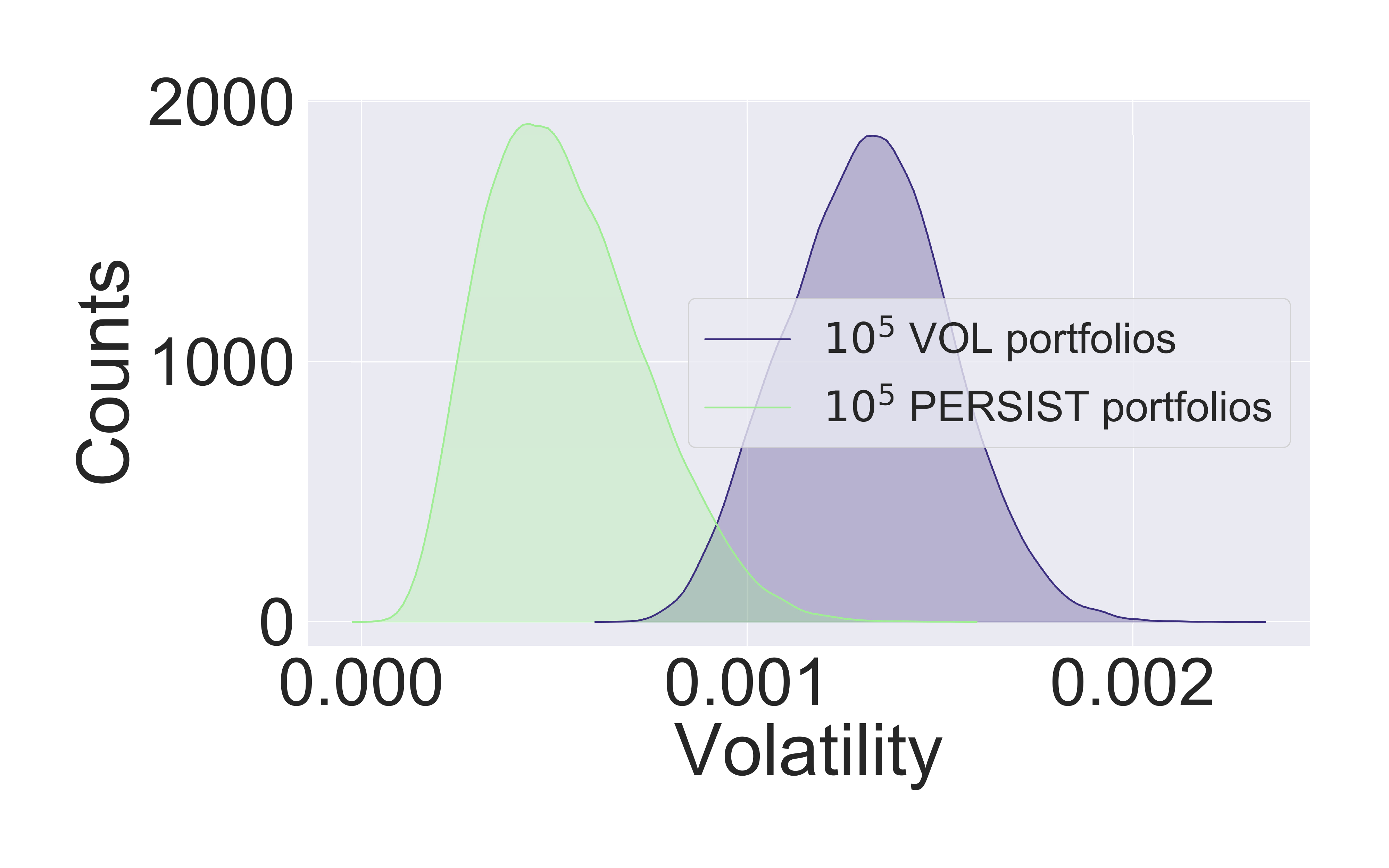}\label{figure10}}
    \subfloat[Israel]{\includegraphics[width=0.49\linewidth]{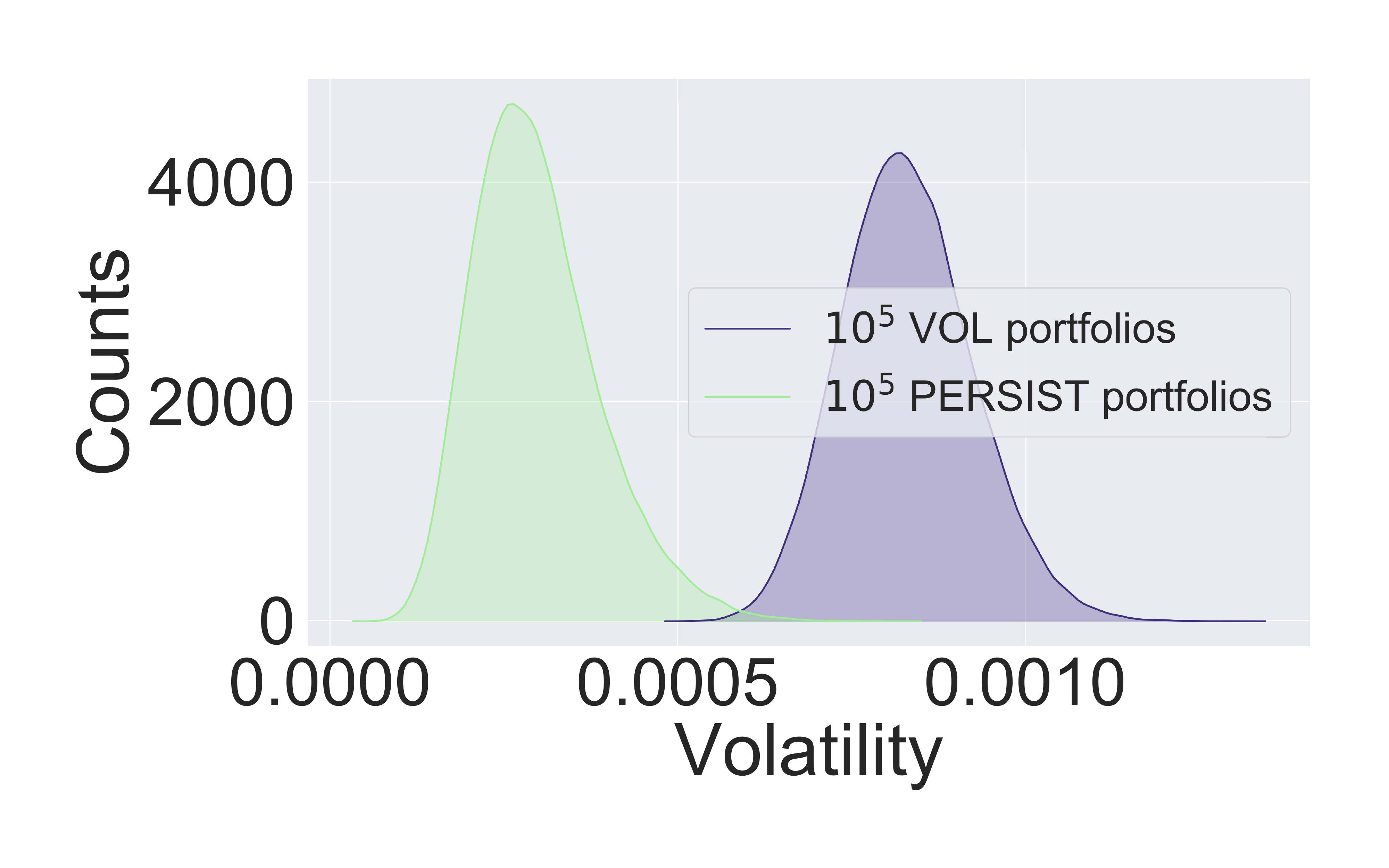}\label{figure11}}
    \caption{Portfolio volatility distribution for the 100 most capitalised stocks in the NYSE (a), German stock market (b), Italian stock market (c) and Israeli stock market (d). The VOL portfolios are formed by weighting a random selection of assets by $1/ \sigma$ while the PERSIST portfolios are formed weighting assets by $1/ P_m(v_i)$.}
    \label{fig_vol_persist_portfolios}
\end{figure}

We now deploy our findings to form a measure directly applicable to portfolio optimisation and compare it with the widespread inverse volatility $1/\sigma$ weighting. In Figure \ref{fig_vol_persist_portfolios} we present out of sample results where we observe a reduction in volatility throughout markets, with distributions separated beyond one standard deviation.

This comparison is meaningful beyond that of an industry standard with a novel approach. The volatility weighting is based on individual assets taken in isolation, weakly influenced by the portfolio's composition of assets. The persistence-based weighting is instead strongly based upon the cliques (therefore the other assets) present in the portfolio. Portfolio composition also influences how the network is filtered, providing a second level of the system's influence on the asset's weighting by persistence. This shows how network analysis and complex systems can greatly enhance our understanding of real world systems beyond traditional methods.

\section{Analysis} \label{analysis}

%\textcolor{red}{NON L'HO ANCORA MODIFICATA DALL'ULTIMA VERSIONE NELL'OTTICA DI GUARDARE PRIMA FINO AI RISULTATI E POI FARE L'ANALISI.}

%We now discuss the results presented in Section \ref{results}.

The power law decay identified in Figure \ref{figure2}, as opposed to an exponential decay,  can be interpreted in terms of long memory in the system. This corroborates observations by Bouchaud et al. and Lillo at al. in \cite{bouchaud2004fluctuations, lillo2004long, bouchaud2009markets, di2005long}, where power law decays in autocorrelation are identified as manifestations of long-memory processes in efficient markets. %Power law decays are meaningful as they are slower than exponential decays, which would represent processes with no long-term memory.

The strict ordering of markets based on their decay exponents in Table \ref{tab1} can be interpreted in terms of more liquid markets being more structured and having less noisy correlations. Suggesting that more efficient and capitalised markets are characterised by structures which are more stable in time. The ordering also leads to the conclusion that more developed markets are characterised by more meaningful underlying structures and cliques, suggesting that systemic risk may represent a greater threat in developed markets. %Furthermore, as per the discussion above, a slower decay indicates stronger long-term memory in the system.
These are interesting observations in relation to the efficient market hypothesis, indicating that more liquid and efficient systems display more stable, autocorrelated and predictable market structures.

The hypothesis of motifs constituting meaningful structures in markets, beyond their individual edges, is strengthened by the results in Table \ref{tab2}. Table \ref{tab2} tests the independence hypothesis of individual edges in motif formation and shows solid evidence to reject it. We can then state that highly persistent motifs are not a mere consequence of highly persistent individual edges, but also of the correlation in those edges existing concurrently.

Table \ref{tab3} demonstrates the need to identify persistent motifs. The ten most persistent motifs visualised in Figure \ref{figure3} are representative of industry sectors in the NYSE. These sectors are not identified by the motifs with higher edge correlation, which instead are dominated by motifs often due to correlation noise in high volatility stocks. Persistent motifs are hence found to be non-trivial with respect to correlation strength of individual edges. The impact on portfolio diversification of the motifs in Figure \ref{figure3} indicates that these structures are highly relevant for diversification in medium to long term investment portfolios. The persistence of these motifs is an intrinsic temporal feature with forecasting power on market structure. As these motifs are not characterised by noticeably strong correlation, a common variance optimisation of the portfolio is unlikely to optimise the weights to sufficiently minimise the effect of these structures. We therefore suggest that filtering of these structures is then perhaps necessary prior to portfolio optimisation, for superior diversification outcomes.

Different results in application to portfolio diversification are presented in Section \ref{portfolio_applications}. We first show how persistent motifs increase the out of sample correlation of a portfolio by comparing portfolios containing all stocks from the ten most persistent motifs in each market with $10^5$ random portfolios with the same number of stocks. We observe the motif portfolio volatility to be significantly above both the mean and median of the random portfolios' volatility distribution. This is a first example of how just selecting stocks from the ten most persistent motifs forms a portfolio with higher long term volatility. Clearly for investment purposes we want the opposite (small volatility). The observation from Figure \ref{fig_random_portfolios} lays the ground for the construction of portfolios where we optimise asset weights in order to reduce the volatility originating from persistent correlations in motif structures.

This is done in Section \ref{portfolio_applications} where we propose a simple node-specific measure for portfolio weighting and selection. We show the out of sample volatility distribution of random portfolios with weights optimised as $1/P_m(v_i)$ to be significantly lower than the distribution of portfolios optimised as $1/\sigma$, a widespread industry standard for portfolio weighting. This result can be explained by the persistence in time being the base of this measure, providing strong out of sample predictive power. Volatility is known to change in the medium to long term for most assets, whilst correlation is also difficult to estimate due to noise in the data and measures. This result greatly enhances the importance and applicability of this work to portfolio optimisation by providing a mapping from persistence-related observations to a direct measure for portfolio optimisation. Future works should investigate a technique to jointly optimise portfolio weighting for both persistence and volatility.

\section{Conclusion} \label{conclusion}

In the present work we have investigated four markets with different capitalisation, liquidity, economic and development characteristics. These markets range from the NYSE to the Israeli Stock Exchange. We construct correlation matrices for 100 stocks from Kendall correlations with exponential smoothing \cite{pozzi2012exponential} and filter them with the TMFG \cite{massara2016network}, as described in Section \ref{method}.

We then base our study on market structure in the form of ``soft'' motif persistence.
A two-regime power law decay in the number of persistent motifs with $\tau$ emerges. The two regimes can be identified via minimisation of the MSE fit measure, providing an unsupervised optimisation method with no tuning parameters. We find that advanced liquid markets are characterised by longer persistence of structured motifs. We argue that this could have consequences for systemic risk.
We discuss long-term memory implications of this decay type and how they allow for forecasting power on market structure. Persistence is studied in order to investigate motif structures and retrieve meaningful sectors in each market. We then show that motif portfolios have significantly higher volatility than random ones. Conclusive results are obtained by defining a node-specific measure for portfolio optimisation and showing that it outperforms volatility-based weighting across markets. This result is of high importance to both practitioners and academics in the context of portfolio optimisation. Future works should investigate a portfolio optimisation technique combining persistence and correlation, based on results and methods from this work. Future works should also investigate applications to portfolio diversification in long-short portfolios and the decay in forecasting and diversification power with time.

\section{Acknowledgments}

TA and JT acknowledge support from the EC Horizon 2020 FIN-Tech project and EPSRC (EP/L015129/1). TA acknowledges support from ESRC (ES/K002309/1); EPSRC (EP/P031730/1) and; EC (H2020-ICT-2018-2 825215).

%
% ---- Bibliography ----
%
%\nocite{*}
\bibliography{ref}

\begin{thebibliography}{10}

\bibitem{markowitz1952portfolio}
Harry Markowitz.
\newblock Portfolio selection.
\newblock {\em The journal of finance}, 7(1):77--91, 1952.

\bibitem{mantegna1999hierarchical}
Rosario~N Mantegna.
\newblock Hierarchical structure in financial markets.
\newblock {\em The European Physical Journal B-Condensed Matter and Complex
  Systems}, 11(1):193--197, 1999.

\bibitem{marcaccioli2019polya}
Riccardo Marcaccioli and Giacomo Livan.
\newblock A p{\'o}lya urn approach to information filtering in complex
  networks.
\newblock {\em Nature communications}, 10(1):745, 2019.

\bibitem{massara2019learning}
Guido~Previde Massara and Tomaso Aste.
\newblock Learning clique forests.
\newblock {\em arXiv preprint arXiv:1905.02266}, 2019.

\bibitem{micciche2019primer}
Salvatore Miccich{\`e} and Rosario~Nunzio Mantegna.
\newblock A primer on statistically validated networks.
\newblock {\em arXiv preprint arXiv:1902.07074}, 2019.

\bibitem{musciotto2018bootstrap}
Federico Musciotto, Luca Marotta, Salvatore Miccich{\`e}, and Rosario~N
  Mantegna.
\newblock Bootstrap validation of links of a minimum spanning tree.
\newblock {\em Physica A: Statistical Mechanics and its Applications},
  512:1032--1043, 2018.

\bibitem{jovanovic2018financial}
Franck Jovanovic, Rosario~N Mantegna, and Christophe Schinckus.
\newblock When financial economics influences physics: The role of
  econophysics.
\newblock {\em Available at SSRN 3294548}, 2018.

\bibitem{cimini2019statistical}
Giulio Cimini, Tiziano Squartini, Fabio Saracco, Diego Garlaschelli, Andrea
  Gabrielli, and Guido Caldarelli.
\newblock The statistical physics of real-world networks.
\newblock {\em Nature Reviews Physics}, 1(1):58, 2019.

\bibitem{kojaku2019constructing}
Sadamori Kojaku and Naoki Masuda.
\newblock Constructing networks by filtering correlation matrices: A null model
  approach.
\newblock {\em arXiv preprint arXiv:1903.10805}, 2019.

\bibitem{masuda2018configuration}
Naoki Masuda, Sadamori Kojaku, and Yukie Sano.
\newblock Configuration model for correlation matrices preserving the node
  strength.
\newblock {\em Physical Review E}, 98(1):012312, 2018.

\bibitem{tumminello2005tool}
Michele Tumminello, Tomaso Aste, Tiziana Di~Matteo, and Rosario~N Mantegna.
\newblock A tool for filtering information in complex systems.
\newblock {\em Proceedings of the National Academy of Sciences},
  102(30):10421--10426, 2005.

\bibitem{massara2016network}
Guido~Previde Massara, Tiziana Di~Matteo, and Tomaso Aste.
\newblock Network filtering for big data: Triangulated maximally filtered
  graph.
\newblock {\em Journal of complex Networks}, 5(2):161--178, 2016.

\bibitem{pozzi2012exponential}
Francesco Pozzi, Tiziana Di~Matteo, and Tomaso Aste.
\newblock Exponential smoothing weighted correlations.
\newblock {\em The European Physical Journal B}, 85(6):175, 2012.

\bibitem{dessi2018supernoder}
Danilo Dess{\`\i}, Jacopo Cirrone, Diego~Reforgiato Recupero, and Dennis
  Shasha.
\newblock Supernoder: a tool to discover over-represented modular structures in
  networks.
\newblock {\em BMC bioinformatics}, 19(1):318, 2018.

\bibitem{musmeci2014risk}
Nicol{\'o} Musmeci, Tomaso Aste, and Tiziana Di~Matteo.
\newblock Risk diversification: a study of persistence with a filtered
  correlation-network approach.
\newblock {\em arXiv preprint arXiv:1410.5621}, 2014.

\bibitem{bouchaud2004fluctuations}
Jean-Philippe Bouchaud, Yuval Gefen, Marc Potters, and Matthieu Wyart.
\newblock Fluctuations and response in financial markets: the subtle nature of
  ‘random’price changes.
\newblock {\em Quantitative finance}, 4(2):176--190, 2004.

\bibitem{lillo2004long}
Fabrizio Lillo and J~Doyne Farmer.
\newblock The long memory of the efficient market.
\newblock {\em Studies in nonlinear dynamics \& econometrics}, 8(3), 2004.

\bibitem{bouchaud2009markets}
Jean-Philippe Bouchaud, J~Doyne Farmer, and Fabrizio Lillo.
\newblock How markets slowly digest changes in supply and demand.
\newblock In {\em Handbook of financial markets: dynamics and evolution}, pages
  57--160. Elsevier, 2009.

\bibitem{di2005long}
Tiziana Di~Matteo, Tomaso Aste, and Michel~M Dacorogna.
\newblock Long-term memories of developed and emerging markets: Using the
  scaling analysis to characterize their stage of development.
\newblock {\em Journal of Banking \& Finance}, 29(4):827--851, 2005.

\end{thebibliography}
\bibliographystyle{unsrt}

\end{document}